\documentclass[11pt, reqno]{amsart}
\usepackage{amsmath, amsthm, a4, latexsym, amssymb}

\def\@rmrk#1#2{\refstepcounter
    {#1}\@ifnextchar[{\@yrmrk{#1}{#2}}{\@xrmrk{#1}{#2}}}

\makeatletter\@addtoreset{equation}{section}\makeatother

 \newfont{\bfit}{cmbxti10 scaled 2000}
 \newfont{\biggi}{cmr12 scaled 2000}

 \newcommand{\eps}{\varepsilon}

 \newcommand{\skrie}{{\mathcal E}}
 
 \newcommand{\skrig}{{\mathcal G}}

 \newcommand{\skril}{{\mathcal L}}
 \newcommand{\skrim}{{\mathcal M}}

 \newcommand{\skrin}{{\mathcal N}}

 \newcommand{\skriy}{{\mathcal Y}}
 \newcommand{\skriz}{{\mathcal Z}}
 
 \newcommand{\sfrac}[2]{\mbox{$\frac{#1}{#2}$}}

\def\1{{\mathchoice {1\mskip-4mu\mathrm l}      
{1\mskip-4mu\mathrm l}
{1\mskip-4.5mu\mathrm l} {1\mskip-5mu\mathrm l}}}

\newcommand{\eq}{\begin{equation}}
\newcommand{\en}{\end{equation}}

{\nopagebreak {\hspace*{\fill}\rule{2mm}{2mm}}\\ }

{\nopagebreak {\hspace*{\fill}\rule{2mm}{2mm}}\\ }

\renewcommand{\subsection}{\secdef \subsct\sbsect}
\newcommand{\subsct}[2][default]{\refstepcounter{subsection}
\vspace{0.15cm}
{\flushleft\bf \arabic{section}.\arabic{subsection}~\bf #1  }
\nopagebreak\nopagebreak}
\newcommand{\sbsect}[1]{\vspace{0.1cm}\noindent
{\bf #1}\vspace{0.1cm}}

\newtheorem{theorem}{Theorem}[section]

\newtheorem{cor}[theorem]{Corollary}

\newtheoremstyle{thm}{1.5ex}{1.5ex}{\itshape\rmfamily}{}
{\bfseries\rmfamily}{}{2ex}{}

\newtheoremstyle{rem}{1.3ex}{1.3ex}{\rmfamily}{}
{\itshape\rmfamily}{}{1.5ex}{}
\theoremstyle{rem}

\refstepcounter{subsection}

\def\thebibliography#1{\section*{References}
  \list%
  {\arabic{enumi}.}
    {\settowidth\labelwidth{[#1]}\leftmargin\labelwidth
    \advance\leftmargin\labelsep
    \parsep0pt\itemsep0pt
    \usecounter{enumi}}
    \def\newblock{\hskip .11em plus .33em minus .07em}
    \sloppy                   
    \sfcode`\.=1000\relax}

\begin{document}
\title[LLLD  for Conditionally Coloured Random  process]
{\Large Local Large  deviations for  empirical locality measure  of   typed Random  Graph Models}

\author[Kwabena Doku-Amponsah]{}

\maketitle
\thispagestyle{empty}
\vspace{-0.5cm}

\centerline{\sc{By Kwabena Doku-Amponsah}}
\renewcommand{\thefootnote}{}
\footnote{\textit{Mathematics Subject Classification :} 94A15,
 94A24, 60F10, 05C80} \footnote{\textit{Keywords: } Local large deviation,conditional  typed  random  graphs, relative  entropy,  method  of types,coupling,  classical  Erdos Renyi graph}
\renewcommand{\thefootnote}{1}
\renewcommand{\thefootnote}{}
\footnote{\textit{Address:} Statistics Department, University of
Ghana, Box LG 115, Legon,Ghana.\,
\textit{E-mail:\,kdoku@ug.edu.gh}.}
\renewcommand{\thefootnote}{1}
\centerline{\textit{University of Ghana}}

\begin{quote}{\small }{\bf Abstract.}
In  this  article,    we  prove a  local  large deviation  principle (LLDP)  for  the   empirical  locality  measure  of   typed random  networks  on $n$ nodes  conditioned to  have  a  given \emph{ empirical type measure}  and \emph{ empirical  link  measure.} From  the LLDP, we  deduce  a full  large  deviation  principle  for  the typed random  graph,  and   the  classical  Erdos-Renyi graphs, where $nc/2$ links are inserted at random among
$n$ nodes. No topological  restrictions  are  required  for  these   results.

\end{quote}\vspace{0.5cm}

\section{Introduction and Background}
\subsection{Introduction.} The  Erdos-Renyi  graph  $\skrig(n,nc/2)$  is  the  simplest  imaginary random,  which  arises  by  taking  $n$  nodes  and  inserting  a fixed  number $nc/2$  links  at  random  among  the  $n$  nodes. See, Van Hofstad~\cite{HDV09}. Some  large deviation  principles  for this  random  graph  have  been  found.  See, Bordenave\& Caputo~\cite{BC13},Doku-Amponsah~\cite{DA06},  and Doku-Amponsah and  Moerters~\cite{DM06a}

Doku-Amponsah and  Moerters~\cite{DM06a}  provided  LDPs  for  the  near-critical  or  sparse  typed  random graphs  with  this model  as a  special  case.  Bordenave  and  Caputo~\cite{BC13}   obtained  large  deviation  principle  for  the  empirical  neigbhourhood  measure  of  the  model  $\skrig(n,nc/2)$.  Large  deviation  for  the  empirical  distribution  was  presented  in  Doku-Amponsah~\cite{DA06} using  the  method of  types and  the  coupling  argument  presented by  Boucheron et. al~\cite{BGL02}.

In  this  paper  we  present  a  Local  Large  deviation  principle  for  the  empirical  locality  measure  of  the  conditional  typed  random  graph  model. See Bakhtin~\cite{BIV15}, Doku-Amponsah~\cite{DA17a}, Doku-Amponsah~\cite{DA17b} for  similar  results  for  the  empirical  measure  of  iid  random  variables,  the  empirical offspring  measure  of  multitype  Galton-Watson  trees  and  the  empirical locality  measure  of  the  sparse  typed random  graphs,   in  which  links  appear  in  the  graph  with  fix  probability  depending  on the  type of  the  links.\\

To  state  our  main  results  in  Section~\ref{AEP} we  need some  notions,  concepts  and  notations from  \cite{DM06a}  and  \cite{DA06a} in  the  background  section  below.

\subsection{Background.}\label{SNS}  Denote  by $\skriz$  a finite alphabet or type set.
Let  $[n]$  be  a fixed set of $n$ nodes, say $[n]=\{1,\ldots,n\}.$
Denote by $\skrig_n$ the set of all (simple) graphs  with node set
$[n]=\{1,\ldots,n\}$ and link set $\skrie.$   By   $\skril(\skriz)$  we  denote  the  set  of all probability  distributions  on  $\skriz,$    $\tilde{\skril}(\skriz)$  we  denote  the  set  of  all  finite   distributions  on  $\skriz,$  and  by  $\skrim(\skriz)$  we  denote  the  set  of  counting  measures  on  $\skriz.$

For any typed graph $Z$( with $n$ nodes), recall   from  \cite{DA06} the definitions  of  the \emph{empirical type distribution}~$P^1\in\skril(\skriz)$, the \emph{empirical link distribution}
$P^2\in\tilde\skril(\skriz\times\skriz)$  and  the \emph{empirical
locality  distribution} $P\in\skril(\skriz\times\skrin(\skriz))$.  For every
$p\in\skril(\skriz\times\skrim(\skriz))$ let $p_1$, $p_2$ be
the $\skriz$-marginal, respectively the  $\skrim(\skriz)-$marginal,
of the measure $p$. Moreover, we define a measure
$\langle p(\cdot,e),\,e(\cdot)\rangle\in\tilde\skril(\skriz\times\skriz)$
by
$$\langle p(\cdot,e),\,e(\cdot)\rangle(a,b):=
\sum_{e\in\skrim(\skriz)}p(a,e)e(b), \quad\mbox{ for
$a,b\in\skriz.$}$$ Define the function  $\Psi \colon
\skril(\skriz\times\skrim(\skriz)) \to \skril(\skriz) \times
\tilde\skril(\skriz \times \skriz)$ by
$\Psi(p)=(p_1,\langle p(\cdot,e),\,e(\cdot)\rangle)$. Note that $\Psi(P)=(P^1, P^2),$ and if these quantities are defined
as empirical locality, type, and link measures of a type
graph.




The  remaining  part  of  the  paper  is  arranged  in  the  following way:  Section~\ref{AEP}  starts  with   statement  of the  main  Theorems  of  the  paper;  Theorem~\ref{smb.tree} and~\ref{smb.tree2}  and Corollary~\ref{smb.tree3}. This results are  proved  in Section~\ref{Proofmain}.  See  subsections~\ref{1},~\ref{2}~and ~\ref{3}.

\section{Statement of main results}\label{AEP}

To present the LDP,  for  a pair of
distributions
$(\pi,p)\in\tilde{\skril}(\skriz\times\skriz)\times\skril(\skriz\times\skrim(\skriz))$  we  recall  from  \cite{DM06a}  the  concept of
\emph{sub-consistency}  and   \emph{consistency} of  measures. We  shall  throughout this  section assume that $\eta(a)>0$ for all  $a\in\skriz.$  We write  $\displaystyle \langle f\,,\,\sigma\rangle:=\sum_{y\in\skriy}\sigma(y)f(y)$  and denote by $$ Q_{(\eta_n, \pi_n)}(z)={\mathbb{Q}}_{(\eta_n, \pi_n)}\{Z=z\}={\mathbb{Q}}\{Z=z | \Psi(P)=(\eta_n, \pi_n)\}$$ the
distribution of the typed  random  graph  $y$ conditioned to
have  typed  law  and  link  law  ${(\eta_n, \pi_n)}$  respectively.
We  define  the  rate  function  $J_{(\eta,\pi)}$  by

\begin{align}\label{randomg.rateLDprob}
J_{(\eta,\pi)}(p)=\left\{
\begin{array}{ll}H(p\,\|\,q) & \mbox
  {if  $(\pi,\,p)$ is  sub-consistent  and $p_1=\eta$  }\\
\infty & \mbox{otherwise.}
\end{array}\right.
\end{align}

where  $$q(a\,,\,e)=\eta(a)\prod_{b\in\skriz}\frac{e^{-\pi(a,b)/\eta(a)}[\pi(a,b)/\eta(a)]^{e(b)}}{e(b)!},\,\mbox{for
$e\in\skrim(\skriz)$} .$$
 The  first  results,  see  Theorem~\ref{smb.tree} below  is  the  local  large  deviation  principle for  typed  random  graphs.
\begin{theorem}[LLDP]\label{smb.tree} Suppose   the  sequence  $(\eta_n, \pi_n)$  converges  to  $(\pi,\eta)\in\skril(\skriz)\times\skril_*(\skriz\times\skriz).$  Let $z=\Big((z(i),\,i\in[n]),\skrie\Big)$ be  coloured  random  graph  conditioned  on  the  event  $\Big\{\Psi(P)=(\eta_n, \pi_n)\Big\}.$  Then,we  have
\begin{itemize}

\item[(i)] for any  functional  $p\in \skril(\skriz\times\skrim(\skriz))$  and a  number  $\eps>0,$  there  exists  a  weak  neighborhood  $B_{p}$  such  that
$$ Q_{(\eta_n, \pi_n)}\Big\{z\in \skrig([n],\, (\eta_n, \pi_n))\,\Big |\, P_z\in B_{p}\Big\}\le e^{-nJ_{(\pi,\,\eta)}(p)-n\eps+o(n)}.$$
\item[(ii)] for  any  $p\in\skril_*(\skriz\times\skriz_*)$, a  number  $\eps>0$  and  a fine  neighborhood  $B_{p}$  we  have  the asymptotic  estimate:
   $$ Q_{(\eta_n, \pi_n)}\Big\{z\in \skrig([n],\, (\eta_n, \pi_n))\,\Big |\, P_z\in B_{p}\Big\}\ge e^{-nJ_{(\pi,\,\eta)}(p)+n\eps-o(n)}.$$
    \end{itemize}
\end{theorem}

We  state  in  Theorem~\ref{smb.tree2}  the  full  LDP  for  the typed  random  graph.

\pagebreak

\begin{theorem}[LDP]\label{smb.tree2}Suppose   the  sequence  $(\eta_n, \pi_n)$  converges  to  $(\pi,\eta)\in\skril(\skriz)\times\skril_*(\skriz\times\skriz).$
 Let $z=\Big((z(i),\,i\in[n]),\skrie\Big)$ be  coloured  random  graph  conditioned  on  the  event  $\Big\{\Psi(P)=(\eta_n, \pi_n)\Big\}.$

\begin{itemize}

\item[(i)]  Let  $F$ be  open subset  of  $ \skril(\skriz\times\skriz_*)$.  Then  we  have
$$\limsup_{n\to\infty}\frac{1}{n}\log Q_{(\eta_n, \pi_n)}\Big\{z\in \skrig([n],\, (\eta_n, \pi_n))\,\Big |\, P_z\in F\Big\}\le - \inf_{p\in F}J_{(\eta, \pi)}(p).$$

\item[(ii)] Let  $\Gamma$ be  closed subset  of  $ \skril(\skriz\times\skriz_*)$. The  we  have

    $$ \liminf_{n\to\infty}\frac{1}{n}\log Q_{(\eta_n, \pi_n)}\Big\{y\in \skrig([n],\, (\eta_n, \pi_n))\,\Big |\, P_z\in \Gamma\Big\}\ge-\inf_{p\in \Gamma}J_{(\eta, \pi)}(p).$$

    \end{itemize}
\end{theorem}
Finally  we  state  from  Theorem~\ref{smb.tree2} a  corollary  for  the  case  of  the  classical  Erdos-Renyi  graph  $\skrig([n],\, nc/2)$.
\begin{cor}[LDP]\label{smb.tree3} Suppose  $D_z$  is  the  degree  distribution  of   $z\in\skrig([n],\, nc/2)$   and  let  $\Gamma $ be  subset  of  $ \skril(\skriz\times\skriz_*)$.  Then  we  have
$$\begin{aligned}
-\inf_{p\in int(\Gamma)}I_{c}(p)&\le\liminf_{n\to\infty}\frac{1}{n}\log Q_{(\eta_n, \pi_n)}\Big\{z\in \skrig([n],\, nc/2)\,\Big |\, D_z\in \Gamma \Big\}\\
&\le\limsup_{n\to\infty}\frac{1}{n}\log Q_{(\eta_n, \pi_n)}\Big\{z\in \skrig([n],\, nc/2)\,\Big |\, D_z\in \Gamma \Big\}\le - \inf_{p\in cl(\Gamma)}I_{c}(p),
\end{aligned}$$
where  $q_c$  is  the  poisson  distribution  with  mean $c$, $I_{c}(p)=H(p\,|\,q_c)$  and  $\infty$ if  otherwise.

\end{cor}

\section{Proof of  Main  Results}\label{Proofmain}


\subsection{Proof  of  Theorem~\ref{smb.tree}.}\label{1}

 We denote by $\tilde{z}$ the multitype  random allocation  process, see  example \cite{DA06} or \cite{DA06a}  and  note that for any   consistent  triple $(\eta_n,\pi_n,p_n)$
we have

\begin{equation}\label{randomg.probcom}
\frac{Q_{(\eta_n, \pi_n)}\big\{Z=z\,\big|P_z\in B_{p}\big\}}{\tilde{Q}_{(\eta_n, \pi_n)}\big\{Z=z\,\big|P_z\in B_{\tilde{p}}\big\}}=\sfrac{\sharp\big\{y:\,P_z=p_n\big\}}{\sharp\big\{z:\,P_z=\tilde{p}_n\big\}}=\sfrac{\sharp\big\{z:\,P_z=p_n\big\}}{\sharp\big\{y:\,(P^1(y),P^2(y))=(\eta_n,\pi_n)\big\}}\times \sfrac{\sharp\big\{z:\,(P^1(z),P^2(z))=(\eta_n,\pi_n)\big\}}{\sharp\big\{z:\,P_z=\tilde{p}_n\big\}}.
\end{equation}

Now  we  define a  locality  of  the  functional  $\rho$  as  follows:
$$B_{p}=\Big\{\mu\in\skril(\skriz\times\skriz_*): \Big \langle \mu, \, \log\sfrac{\mu}{q}\Big\rangle>\Big \langle p, \,\log\sfrac{p}{ q}\Big \rangle -\sfrac{\eps}{2}\Big\}.$$

Using  Doku-Amponsah~\cite{DA06},  under  the  condition  $P_z\in B_{p}$  and  $\tilde{p}=q$   we  have  that

\begin{equation}\label{Equ2}
e^{-nH(p_n\,\|\,q_n)+nH(q\,\|\,q_n)-o(n)} \le
\frac{dQ_{(\eta_n, \pi_n)}(z)}{d\tilde{Q}_{(\eta_n, \pi_n)}(z)}\le
e^{-nH(p_n\,\|\,q_n)+nH(q\,\|\,q_n)+o(n)},\end{equation}

Fix  $\eps>0,$   and  note  that  by  \ref{Equ2} we  have

$$ \begin{aligned}
Q_{(\eta_n, \pi_n)}\Big\{z\in \skrig([n],\, (\eta_n, \pi_n))\,\Big |\, P_z\in B_{p}\Big\}&\le\int_{P_z\in B_{p}} e^{-nH(p_n\,\|\,q_n)+nH(q\,\|\,q_n)+o(n)}d\tilde{Q}_{(\eta_n, \pi_n)}(z)\\
&  \le e^{-nJ_{(\eta, \pi)}(p)-n\eps+o(n)}.
\end{aligned}$$

\pagebreak

Also  we  have
$$\begin{aligned}
Q_{(\eta_n, \pi_n)}\Big\{z\in \skrig([n],\, (\eta_n, \pi_n))\,\Big |\, P_z\in B_{p}\Big\}&\ge\int_{P_z\in B_{p}} e^{-nH(p_n\,\|\,q_n)+nH(q\,\|\,q_n)-o(n)}d\tilde{Q}_{(\eta_n, \pi_n)}(z) \\
&\ge e^{-nJ_{(\eta, \pi)}(p)-n\eps-o(n)},
\end{aligned}$$
which  completes  the  proof  of  the  Theorem.  The  proof  of Theorem~\ref{smb.tree2}   below,  follows  from  Theorem~\ref{smb.tree} above  using  similar  arguments  as  in \cite[p. 544]{BIV15}.
\subsection{Proof  of  Theorem~\ref{smb.tree2}.}\label{2}
\begin{proof}
 We observe  that  the  empirical  locality  distribution  is  a  probability  distribution  and  so it  is contain in  the  unit  ball  in  $\tilde{\skril}(\skriz\times\skriz^*).$ Henceforth,  without  loss  of  generality  we can  assume  that  the  set  $\Gamma$  in  Theorem~\ref{smb.tree2}(ii)   above  is  relatively  compact.   If  we  choose  any  $\eps>0,$   then  for  each  functional  $p\in \Gamma$  we can  find  a  weak  neigbourhood  such  that the  estimate  of  Theorem~\ref{smb.tree}(i)  above  holds.  From  all  these  neigbhoourhood,  we  choose a  finite  cover  of  $\skrig([n],\, (\eta_n, \pi_n))$  and  sum up over  the  estimate  in  Theorem~\ref{smb.tree}(i) above  to  obtain

$$ \lim_{n\to\infty}\frac{1}{n}\log Q_{(\eta_n, \pi_n)}\Big\{z\in \skrig([n],\, (\eta_n, \pi_n))\,\Big |\, P_z\in \Gamma\Big\}\le-\inf_{p\in \Gamma}J_{(\eta, \pi)}(p)+\eps.$$
 Since $\eps$ was  arbitrarily  chosen  and  the   lower  bound  in  Theorem~\ref{smb.tree}(ii)  is implies  the  lower  bound  in  Theorem~\ref{smb.tree2}(i) we  have  the  required  results  which  completes the  proof.

\end{proof}

\subsection{Proof  of  Corollary~\ref{smb.tree3}.}\label{3}
\begin{proof}
The  proof  of  Corollary~\ref{smb.tree3}  follows  from  Theorem~\ref{smb.tree2}  if  we  take  $P=D_z$  then $\Psi(D_z)=2|\skrie|/n=c.$  Hence, in  the  case  of  the classical  Erdos-Renyi  graph  $\skrig([n],\,nc/2)$, we  have  the convex rate  function
 $$
I_c(p)=J_{(\eta,\, \pi)}(p)$$    which reduces  to   $H(p\,|\,q_c)$  if  the  probability measure $p$  has  mean $c$  and  $\infty$  if  otherwise.  
\end{proof}

{\bf Acknowledgement}

\emph{This  article  was  finalized  at  the  Carnigie Banga-Africa, June 27-July  2017  writeshop, in  Koforidua.}




\begin{thebibliography}{WWW98}


\bibitem[1]{BC13}
{\sc C.~Bordenave   and  P.~Caputo}
\newblock{Large  deviations  of  empirical neighbourhood  distribution in sparse random graphs.}
\newblock{\emph{arxiv:1308.5725(2013).}}
\smallskip


\bibitem[2]{BGL02}
 {\sc S. Boucheron,} {\sc F. Gamboa }and {\sc C. Leonard.}
 \newblock{ Bins and balls: Large deviations of the empirical occupancy process.}
\newblock{\emph{Ann. Appl. Probab. 12 607-636 (2002).}}
\smallskip

\bibitem[3]{BIV15}
{\sc I.V.~Bakhtin.}
\newblock{Spectral Potential,  Kullback  Action,  and  Large  deviations  of  empirical measureson  measureable  spaces.}
\newblock{\emph{Theory  of  Probability  and  application. Vol.  50,No.4.(2015) pp.535-544.}}
\smallskip

\bibitem[4]{DA06a}
{\sc K.~Doku-Amponsah.}
\newblock{Large  deviations and basic information theory for  hierarchical  and networked data structures.}
\newblock {\emph{Bath Thesis(2006).}}
\smallskip

\bibitem[5]{DA06}
{\sc K.~Doku-Amponsah.}
\newblock{Exponential Approximation  of  empirical neigbourhood  measures  of  random  graphs by  random  allocation.}
\newblock {\emph{International J. of Statistics and Probability, Vol. 3, No 2; 2014. pp 110-120.}}
\smallskip

\bibitem[6]{DA17a}
{\sc K.~Doku-Amponsah.}
\newblock{ Local Large  deviations: MCMillian Theorem for  multitype Galton-Watson  Processes.}
\newblock {\emph{Far East  J. of Mathematical  Sciences 102(10),pp. 313-337}}
\smallskip

\bibitem[7]{DA17b}
{\sc K.~Doku-Amponsah.}
\newblock{Local Large  deviations: a  McMillian Theorem for  Typed Random Graph  Processes.}
\newblock {\emph{ J. of  Mathematics  and  Statistics 13(4), pp.325-329}}
\smallskip

\bibitem[8]{DM06a}
{\sc K.~Doku-Amponsah}  and {\sc P.~M\"orters.}
\newblock{Large deviation principle for  empirical measures of
coloured random graphs.}
\newblock { \emph{The  Annals  of  Applied  Probability, 20(6), 1989-2021.}}

\bibitem[9]{DZ98}
{\sc A.~Dembo} and {\sc O.~Zeitouni.}
\newblock Large deviations techniques and applications.
\newblock Springer, New York, (1998).
\smallskip


\bibitem[10]{HDV09}
{\sc R. Van  Der  Hofstad}
\newblock {Random  Graphs and  Complex  Networks.}
\newblock {\emph{Eindoven University  of  Technology.} Unpublished  Manuscript.}
\smallskip

\end{thebibliography}
\end{document}